\documentstyle[12pt]{article}
\newlength{\mathspace}
\def\b #1{\bar{#1}}
\def\np#1{ Nucl. Phys. B#1}
\def\pr#1    { Phys. Rev. D#1 }
\def\pl#1{ Phys. Lett. B#1}

\def\ijmp#1  { Int. Jour. Mod. Phys. A#1 }
\def\mpl#1   { Mod. Phys. Lett. A#1 }
\def\begineq{\begin{equation}}
\def\endeq{\end{equation}}
\def\eqabegin{\begin{eqnarray}}
\def\eqaend{\end{eqnarray}}
\def\nn{\nonumber}

\begin{document}
\baselineskip=0.7cm
\setlength{\mathspace}{2.5mm}



\begin{titlepage}

    \begin{normalsize}
     \begin{flushright}
                 CERN-TH/97-225, SINP-TNP/97-11\\
                 hep-th/9709017\\
     \end{flushright}
    \end{normalsize}
    \begin{LARGE}
       \vspace{5mm}
       \begin{center}
         {An SL(2, Z) Multiplet of Black Holes}\\ 
         {in $D = 4$ Type II Superstring Theory}\\ 
       \end{center}
    \end{LARGE}

  \vspace{5mm}

\begin{center}
           
             \vspace{.5cm}

            Ashok D{\sc as}

\vspace{1mm}

{\it Department of Physics and Astronomy}\\
{\it University of Rochester, Rochester, NY 14627, USA}

\vspace{1mm}

Jnanadeva M{\sc aharana}\footnote[3]{Jawaharlal Nehru Fellow}
\footnote[2]{E-mail address: maharana@nxth04.cern.ch}

\vspace{1mm}

{\it CERN, Geneva, Switzerland}

\vspace{1mm}

and

\vspace{1mm}

{\it Institute of Physics,
 Bhubaneswar 751 005, India}

\vspace{2mm}

and

\vspace{2mm}

            Shibaji R{\sc oy}
           \footnote{E-mail address: roy@tnp.saha.ernet.in} 

                 \vspace{1mm}

{\it Saha Institute of Nuclear Physics}\\
        {\it 1/AF Bidhannagar, Calcutta 700 064, India}\\

      \vspace{5mm}

    \begin{large} ABSTRACT \end{large}
        \par
\end{center}
 \begin{normalsize}
\ \ \ \
It is well-known that the conjectured SL(2, Z) invariance of type IIB
string theory in ten dimensions also persists in lower dimensions when
the theory is compactified on tori. By making use of this recent observation,
we construct an infinite family of magnetically charged black hole solutions
of type II superstring theory in four space-time dimensions. These solutions
are characterized by two relatively prime integers corresponding to the
magnetic charges associated with the two gauge fields (from NS-NS and R-R
sectors) of the theory and form an SL(2, Z) multiplet. In the extremal limit
these solutions are  stable as they are prevented from decaying
into black holes of lower masses by a `mass gap' equation.
\end{normalsize}

\end{titlepage}
\vfil\eject



Black hole solutions in string theory [1] provide a very interesting
arena to address some of the long-standing issues involving 
thermodynamics of black holes, their 
evaporation and information loss paradox[2]. It is well-known that string
theories admit a rich variety of static spherically symmetric as
well as rotating black hole solutions in various dimensions.
For example, spherically symmetric black hole solutions in string
theory having purely magnetic, purely electric and both charges
(dyonic black holes) have been constructed before [3--6]. Furthermore,  rotating
black hole solutions containing electric, magnetic and both
charges have been discussed in refs.[7,8]. Since black holes have
profound conceptual implications in our understanding of the
nature of general relativity in the quantum domain and since
string theory is believed to lead to a finite, consistent theory
of quantum gravity, it is very important to construct various
kinds of black hole solutions in string theory and study their
properties. As black holes are intrinsically non-perturbative, it
is in general difficult to study their properties in the perturbative
framework of string theory. However, there has been a 
spectacular  advancement
in  our understanding of the non-perturbative behavior of string
theory in recent times. Subsequently,  Strominger and Vafa [9] 
constructed a special class of
black hole solutions in type II string theory in $D = 5$ and
reproduced the Bekenstein-Hawking area entropy relation through
a D-brane description [10] of such black holes and by 
counting the number
of microstates in this framework. Black holes in that case
saturate the BPS condition in the extremal limit and carry an
electric as well as an axionic charge. Further developments along
this line could be found in refs.\footnote[2]{These contain only a partial
list.}[11,12].

Recently, 
we have shown [13,14] that the low energy effective action of type IIB
string theory has a manifest SL(2, R) invariance in lower
dimensions when compactified on tori as a consequence of the
corresponding symmetry in ten dimensions [15--18]. This symmetry is
non-perturbative as it transforms the string coupling constant
in a non-trivial way. A discrete subgroup of this SL(2, R)
group has been conjectured to be an exact symmetry of the
quantum type IIB string theory. A strong evidence in favor of
this conjecture has been given in ref.[19], when we showed that
there exist SL(2, Z) multiplets of macroscopic string-like 
solutions in type II string theories in $D < 10$. The tensions
and the charges of these BPS saturated string-like solutions
have been shown to be given by SL(2, Z) covariant expressions.

In this paper, we will construct another class of black hole
solution in $D = 4$ type II string theory. 
By making use of this SL(2, R) invariance of the lower dimensional
type II string theory, we construct an SL(2, Z) multiplet
of black hole solutions in $D = 4$. First, we construct a magnetically
charged black hole solution, similar to the one obtained by Garfinkle,
Horowitz and Strominger (GHS) [4]. This solution arises due to the presence 
of an Abelian gauge field in the NS-NS sector of the theory in four dimensions
as a consequence of compactification of the corresponding antisymmetric tensor
field in the ten dimensional action. Next, we implement the SL(2, R)
transformations so that the resulting solution carries both NS-NS and R-R
charges\footnote[2]{In  the recent discussions on the microscopic
origin of the Bekenstein-Hawking area entropy relation [9,11,12], black
holes with R-R charges have been considered and this is crucial to
have D-brane description of the black holes.}. 
 The two Abelian gauge fields 
 correspond to the  dimensionally reduced 
antisymmetric tensor fields coming from the NS-NS and R-R sectors
of type II string theory.
As the magnetic charges are quantized, the final solution
will, therefore, be characterized by the two integers corresponding
to the magnetic charges associated with the two gauge fields
of NS-NS and R-R sectors. In the extremal case, we will show that
both the charges and the masses of such black holes are given by
SL(2, Z) covariant expressions. Since in the extremal limit the
magnetically charged black holes are BPS saturated, the SL(2, Z)
covariant results give a strong evidence in favor of the conjectured
SL(2, Z) invariance of the quantum theory. We mention in passing  
 that type II string theory in four dimensions has been
conjectured to possess a much bigger non-compact global 
symmetry group E$_{7(7)}$(Z) [20,16],
known as the U-duality group [16], which contains both the S-duality
[21] group SL(2, Z) and the T-duality [22] group O(6, 6; Z) 
as the subgroup.
But we will restrict ourselves only to a part of this bigger symmetry
group, namely, the S-duality group. Then we will show that these
extremal black holes are stable when they are characterized
by two relatively prime integers. In that case, as common to
BPS saturated states, the masses of the black holes satisfy a
triangle inequality which prevents the black holes to decay into 
black holes of lower masses.





Let us recapitulate how the four dimensional magnetically charged black hole
solution of GHS [4] arises from four dimensional effective action.

The complete low energy four dimensional effective action  of interest to us is 
\eqabegin
S &=& \int\,d^4 x \sqrt{-G} e^{-2\phi} \left(R + 4 \partial_\mu
\phi \partial^\mu \phi + \frac{1}{8} {\rm tr}\,\partial_\mu {\bf M}^{-1}
\partial^\mu {\bf M} \right.\nn\\
& &\qquad\qquad\qquad\qquad \left. 
-\frac{1}{12} H_{\mu\nu\rho} H^{\mu\nu\rho}
-\frac{1}{4} {\cal F}_{\mu\nu}^T {\bf M}^{-1} {\cal F}^{\mu\nu}\right)
\eqaend
where $G = ({\rm det}\,G_{\mu\nu})$, $G_{\mu\nu}$ being the four
dimensional metric in the string frame, 
$\phi$ is the dilaton field in $D = 4$, $R$
is the scalar curvature corresponding to the  metric $G_{\mu\nu}$. 
This four dimensional action is of generic form which can be obtained through
toroidal compactification on $T^6$ of a ten dimensional string effective 
action. For example, if we start from the ten dimensional  heterotic string,
the matrix ${\bf M}$ which contains the scalar fields, 
parametrizes the coset, $
{O(22,6)} \over { O(22) \times O(6)} $ and ${\cal F}_{\mu \nu}$ corresponds to
28 Abelian gauge field strengths [23]. On the other hand if we start from 
ten dimensional action of type II theories, then the reduced action (1) 
can be identified with the one that is obtained by dimensional reductions
of the NS-NS sector and now there will be only 12 gauge fields (6 from the
metric and 6 from antisymmetric tensor) and ${\bf M}$ will contain scalars
parametrizing the coset $\frac{O(6,6)}{O(6) \times O(6)}$.
 The superscript `$T$' denotes the transpose of a matrix. Definitions of the
field strengths are 
\eqabegin
{\cal F}_{\mu\nu} &=& \partial_\mu {\cal A}_\nu - \partial_\nu {\cal
A}_\mu\nn\\
H_{\mu\nu\rho} &=& \partial_\mu B_{\nu\rho} + {\cal A}_\mu^T \eta 
{\cal F}_{\nu\rho} + {\rm cyc.\,\,in\,\,\mu\nu\rho}
\eqaend
where ${\cal A}_\mu$ is a 28 dimensional vector field containing the
28 gauge fields coming from the dimensional reduction of the ten
dimensional metric, antisymmetric tensor field and U(1)$^{16}$ gauge
fields in the case of heterotic string.

In order to obtain magnetically charged black hole solution, we choose
${\bf M}$ to be constant and put   $H_{\mu\nu
\rho} = 0$ and set all the gauge fields except 
one (denoted as $A_\mu^{(1)}$)
to zero, then the action (1) reduces in the Einstein frame to,
\begineq
{\bar S} = \int\, d^4 x \sqrt{-g} \left(R - 2\partial_\mu \phi
\partial^\mu \phi - \frac{1}{4} e^{-2\phi} F_{\mu\nu}^{(1)} 
F^{(1)\,\mu\nu}\right)
\endeq
where the Einstein metric is related to the string metric by $g_{\mu\nu}
= e^{-2\phi} G_{\mu\nu}$. $R$ now denotes the scalar curvature with
respect to the Einstein metric $g_{\mu\nu}$. Note that this action is
precisely the one considered by GHS [4]. In their case, the gauge field 
$A_\mu^{(1)}$ came from one of the U(1)$^{16}$ gauge fields in ten
dimensions, whereas, we choose $A_\mu^{(1)}$ to come from the dimensional
reduction of the ten dimensional antisymmetric tensor field. 

Our conventions and notations are as follows: the signature of the tangent 
space Lorentz metric is ($-,\,+,\,+,\,\ldots$). 
The covariant derivative, connection and the Riemann curvature
tensor are:
\eqabegin
\nabla_\mu V_\nu 
&=& \partial_\mu V_\nu - \Gamma^\lambda_{\,\,\,\,\mu\nu} V_
\lambda\nn\\
\Gamma^\mu_{\,\,\,\,\nu\lambda} &=& \frac{1}{2} g^{\mu\rho}\left(
\partial_\lambda g_{\rho\nu} + \partial_\nu g_{\rho\lambda} - \partial_\rho
g_{\nu\lambda}\right)\\
R^\mu_{\,\,\,\,\nu\lambda\rho} &=& \partial_\lambda \Gamma^\mu_{\,\,\,\,
\nu\rho} - \partial_\rho \Gamma^\mu_{\,\,\,\,\nu\lambda} + \Gamma^\mu_{
\,\,\,\,\lambda\sigma} \Gamma^\sigma_{\,\,\,\,\nu\rho} - \Gamma^\mu_{
\,\,\,\,\rho\sigma} \Gamma^\sigma_{\,\,\,\,\nu\lambda}\nn
\eqaend
where $V_\mu$ is any vector. The scalar curvature is given as $R = 
g^{\mu\nu} R_{\mu\nu} = g^{\mu\nu} R^\rho_{\,\,\,\,\mu\rho\nu}$.

The equations of motion derived from the action (3) are as given by
\eqabegin
& &\nabla_\mu\left(e^{-2\phi} F^{(1)\,\mu\nu}\right) \,\,\,=\,\,\,0\\
& &\nabla^2 \phi + \frac{1}{8} e^{-2\phi} F_{\mu\nu}^{(1)} F^{(1)\,\mu\nu}
\,\,\,=\,\,\,0\\
& &R_{\mu\nu}\,\,\,=\,\,\,2\partial_\mu \phi \partial_\nu \phi + \frac{1}
{2} e^{-2\phi} F_{\mu\lambda}^{(1)} F_\nu^{(1)\,\,\,\lambda} - \frac{1}{8}
g_{\mu\nu} e^{-2\phi} F_{\lambda\rho}^{(1)} F^{(1)\,\lambda\rho}
\eqaend
A static spherically symmetric black hole solution of these equations
can be obtained from the following ansatz of the space-time metric
\eqabegin
ds^2 &=& -f^2 dt^2 + f^{-2} dr^2 + R^2 d\Omega\nn\\
&=& -f^2 dt^2 + f^{-2} dr^2 + R^2\left(d\theta^2 + \sin^2 \theta d
\varphi^2\right)
\eqaend
and the Maxwell field
\begineq
F_{23}^{(1)} = Q \sin\theta
\endeq
We denote the coordinates $t$, $r$, $\theta$ and $\varphi$ by 0, 1, 2
and 3 respectively. $f$,  $R$  and $\phi$ are functions of the radial coordinate
$r$ only. Asymptotically, as 
$r \rightarrow \infty$, $f \rightarrow 1$, $R \rightarrow r$ and $\phi
\rightarrow \phi_0$, otherwise the functions are arbitrary. We also note 
from (9) that since $F_{23}^{(1)}$ is the only non-zero component of the
Maxwell field, it is magnetic and $Q$ is the corresponding charge defined
as $Q \equiv \frac{1}{4\pi} \int F^{(1)}$. It can be easily checked
from (8) and (9) that $F_{\mu\nu}^{(1)} F^{(1)\,\mu\nu} = 2 Q^2/R^4$.
Also, the only non-zero components of the connection we find from 
(4) and (8) are
\eqabegin
\Gamma^0_{\,\,\,\,01} &=& \frac{f'}{f}\nn\\
\Gamma^1_{\,\,\,\,00} &=& f^3 f',\quad \Gamma^1_{\,\,\,\,11}\,\,\,=\,\,\,
-\frac{f'}{f},\quad \Gamma^1_{\,\,\,\,22}\,\,\,=\,\,\,-f^2 R R',\quad
\Gamma^1_{\,\,\,\,33}\,\,\,=\,\,\,-f^2 R R' \sin^2\theta\nn\\
\Gamma^2_{\,\,\,\,33} &=& -\sin\theta \cos\theta,\quad \Gamma^2_{\,\,\,\,
12}\,\,\,=\,\,\,\frac{R'}{R}\\
\Gamma^3_{\,\,\,\,13} &=& \frac{R'}{R}\quad {\rm and} \quad \Gamma^3_{\,\,
\,\,23}\,\,\,=\,\,\,\cot\theta\nn
\eqaend
Here `prime' denotes the derivative with respect to the radial coordinate
$r$. Note that  Eq.(5) is automatically
satisfied using above information and then   from Eq.(7) we arrive at the 
relations,
\eqabegin
R_{00} &=& \frac{f^2 Q^2}{4 R^4} e^{-2\phi}\,\,\,=\,\,\, -f^2
\nabla^2 \phi\\
R_{22} &=& \frac{Q^2}{4 R^2} e^{-2\phi}\,\,\,=\,\,\, -R^2 \nabla^2\phi
\eqaend
where we have made use of Eq.(6) to write the last expressions in (11) and
(12). By  comparing (11) and (12) we have,
\begineq
R_{00} = \frac{f^2}{R^2} R_{22}
\endeq
Furthermore,  $R_{00}$ and $R_{22}$ can also be expressed in terms of the 
functions $f$ and $R$, appearing in the metric (8), through the
definition of  Ricci tensor (4) as
follows,
\eqabegin
R_{00} &=& f^2\left(ff'' + (f')^2 + 2ff'\frac{R'}{R}\right)\\
R_{22} &=& 1 - \left(f^2 R R'\right)'
\eqaend
where we have made use of (10). Substituting (14) and (15) in (13),
we obtain an equation involving the two unknown functions $f$ and
$R$ as,
\begineq
\left(f^2 R^2\right)'' = 2
\endeq
Now using (11), (14) and the expression for $\nabla^2 \phi = f^2 \phi''
+ 2 f f' \phi' + 2 f^2 \frac{R'}{R} \phi'$, we obtain another equation
involving $f$, $R$ and the dilaton $\phi$ as,
\begineq
\left(\frac{f^2 R^2 X'}{X}\right)' = 0
\endeq
where,   $X \equiv f^2 e^{2\phi}$. Eqs.(16) and (17)
can now be easily solved if we  impose  the asymptotic limit
as $r \rightarrow \infty$, $f \rightarrow 1$ and $R \rightarrow r$.
The solution is:
\eqabegin
f^2(r) &=& \left(1 + \frac{a}{r}\right)\nn\\
R^2(r) &=& r^2\left(1 + \frac{b}{r}\right)\\
e^{-2\phi} &=& e^{-2\phi_0} \left(1 + \frac{b}{r}\right)\nn
\eqaend
with $ab = \frac{1}{2} Q^2 e^{-2\phi_0}$. Here $a$, $b$ are integration
constants and to find $ab$, we have used Eq.(6). The integration constant
`$a$' can be identified from the weak field limit\footnote[2]{We have 
set the Newton's constant $G =1$.} as $-2M$, where $M$ is the mass of
the black hole and therefore $b = - \frac{Q^2}{4M} e^{-2\phi_0}$, $Q$
being the magnetic charge of the black hole defined earlier. The 
background field configurations, therefore, take the following form:
\eqabegin
ds^2 &=& -\left(1 - \frac{2M}{r}\right) dt^2 + \left(1 - \frac{2M}{r}\right)
^{-1} dr^2 + r^2\left(1 - \frac{Q^2}{4Mr}e^{-2\phi_0}\right) d\Omega\\
e^{-2\phi} &=& e^{-2\phi_0}\left(1 - \frac{Q^2}{4Mr} e^{-2\phi_0}\right)
\,\,\,=\,\,\,e^{-2\phi_0}\frac{R^2}{r^2}\,\,\,=\,\,\,e^{-2\phi_0}
\frac{(1-f^2)^2 R^2}{4M^2}\\
F_{23}^{(1)} &=& Q\sin\theta
\eqaend
The solution may be written in a more symmetric fashion by introducing
the dilaton charge ${\cal D} \equiv \frac{1}{4\pi}\int\,d^2\Sigma^\mu 
\nabla_\mu\phi
= \frac{R^2(r)}{4\pi}\int\,d\Omega \phi' = - \frac{Q^2}{8M} e^{-2\phi_0}$
as,
\eqabegin
ds^2 &=& -\left(1 - \frac{2M}{r}\right) dt^2 + \left(1 - \frac{2M}{r}\right)
^{-1} dr^2 + r^2\left(1 - \frac{2|{\cal D}|}{r}\right) d\Omega\nn\\
e^{-2\phi} &=& e^{-2\phi_0} \left(1 - \frac{2|{\cal D}|}{r}\right)
\eqaend
Now with the coordinate transformation of the form:
\eqabegin
\rho^2 &=& r^2 \left(1 - \frac{2|{\cal D}|}{r}\right), 
\quad {\rm for} \quad
r \geq 2|{\cal D}|\nn\\
&=& -r^2 \left(1 - \frac{2|{\cal D}|}{r}\right),
\quad {\rm for} \quad r \leq
2|{\cal D}|
\eqaend
which implies, $r = |{\cal D}| + \sqrt{{\cal D}^2 + \rho^2}$, for 
$r \geq 2|{\cal D}|$, 
$r = |{\cal D}| + \sqrt{{\cal D}^2 - \rho^2}$, for $|{\cal D}| 
\leq r \leq 2|{\cal D}|$ and
$r = |{\cal D}| - \sqrt{{\cal D}^2 - \rho^2}$, for $r\leq 
|{\cal D}|$, it
can be easily checked that (22) represents a black hole with an event 
horizon located at
\begineq
\rho = 2 \left[M\left(M - |{\cal D}|\right)\right]^{\frac{1}{2}}
\endeq
for $M > |{\cal D}|$. However, for $M < |{\cal D}|$, 
there is no event horizon and 
consequently the space-time singularity at $r = 0$ is directly observable
representing a ``naked'' singularity. Thus the transition between the 
black hole and the ``naked'' singularity occurs at $M = |{\cal D}| 
= \frac{Q^2}
{8M} e^{-2\phi_0}$ or $Q^2 = 8M^2 e^{2\phi_0}$. The transition point is
known as the extremal limit. Note from (22) that at the extremal limit
the area of the event horizon vanishes causing the surface to be singular.
Thus  the  black hole solution of GHS with magnetic charge was obtained 
for  a special background configuration of 
four dimensional heterotic string theory  and the one we presented 
exactly coincides with GHS; however, the four dimensional action is the 
NS-NS sector of type II theory as remarked earlier. Now, we proceed to
discuss compactification of type IIB theory to four dimensions with
relevant massless fields in both NS-NS and R-R sector.




Let us recall that the massless spectrum of the type IIB string
theory in the bosonic sector contains a graviton, a dilaton and
an antisymmetric tensor field as NS-NS sector states, whereas, in
the R-R sector it contains another scalar, another antisymmetric
tensor field and a four-form gauge field whose field-strength
is self-dual. It is well known that a covariant action for self dual five 
index antisymmetric tensor fields in ten dimensions does not exist [24]
and  we set this field strength to zero, since this field is of 
no relevance to us
in what follows. Therefore, a 
consistent, covariant action can be written [17] from which the type IIB
supergravity equations of motion can be derived. We have studied
the dimensional reduction of this action on a $(10 - D)$ dimensional
torus in ref.[13,14]. When $D = 4$, the corresponding four dimensional
type IIB string effective action in the Einstein frame takes the
following form:
\eqabegin
\b {S}_{\rm II}&=&\int\,d^4 x \sqrt{-g}\left[R + \frac{1}{4} 
{\rm tr}\,
\partial_\mu {\cal M} \partial^\mu {\cal M}^{-1} +
\frac{1}{8} \partial_\mu \log
\b {\Delta} \partial^\mu \log \b {\Delta} + \frac{1}{4} \partial_\mu
g_{mn} \partial^\mu g^{mn}\right.\nn\\
& &\qquad\qquad\qquad -\frac{1}{4} g_{mn} F_{\mu\nu}^{(3)\,m} F^{(3)\,
\mu\nu,\,n} - \frac{1}{4} (\b \Delta)^{1/2} g^{mp} g^{nq}
\partial_\mu {\cal B}_{mn}^T {\cal M} \partial^\mu {\cal B}_{pq}\\
& &\qquad\qquad\qquad\left. -\frac{1}{4} 
(\b \Delta)^{1/2} g^{mp} {\cal H}
^T_{\mu\nu\,m} {\cal M} {\cal H}^{\mu\nu}_{\,\,\,\,p} - \frac{1}{12}
(\b \Delta)^{1/2} {\cal H}_{\mu\nu\lambda}^T {\cal M} 
{\cal H }^{\mu\nu\lambda}\right]\nn
\eqaend
Here $g = ({\rm det}\,g_{\mu\nu})$, where $g_{\mu\nu}$ is the
four dimensional Einstein metric and 
 $R$ is the scalar
curvature associated with $g_{\mu\nu}$. ${\cal M}$ is an SL(2, R)
matrix defined as
\begineq
{\cal M} \equiv \left(\begin{array}{cc} \chi^2 + e^{-2\tilde {\phi}} &
\chi \\ \chi & 1\end{array}\right) e^{\tilde {\phi}}
\endeq
where $\chi$ is the R-R scalar and $\tilde {\phi} = \phi + \frac{1}{2}
\log \Delta$, $\phi$ being the NS-NS scalar, the four dimensional dilaton
and $\Delta^2 = ({\rm det}\, G_{mn})$, $G_{mn}$ being the scalars coming 
from the dimensional reduction of the ten dimensional string metric.
$g_{mn} = e^{-2\phi} G_{mn}$ and $(\b {\Delta})^2 = ({\rm det}\,g_{mn})$.
$F_{\mu\nu}^{(3)\,m} = \partial_\mu A_\nu^{(3)\,m} - \partial_\nu A_\mu
^{(3)\,m}$, where $A_\mu^{(3)\,m}$ is the gauge field resulting from the
dimensional reduction of the string metric. ${\cal B}_{mn} \equiv \left(
\begin{array}{c} B_{mn}^{(1)} \\ B_{mn}^{(2)}\end{array}\right)$, where
$B_{mn}^{(i)}$, for $i = 1, 2$ are the moduli coming from the dimensional
reduction of the NS-NS and R-R antisymmetric tensor fields. ${\cal H}_
{\mu\nu\,m} \equiv \left(\begin{array}{c} H_{\mu\nu\,m}^{(1)} \\
H_{\mu\nu\,m}^{(2)}\end{array}\right)$, where $H_{\mu\nu\,m}^{(i)} =
F_{\mu\nu\,m}^{(i)} - B_{mn}^{(i)} F_{\mu\nu}^{(3)\,n}$ and 
$F_{\mu\nu\,m}^{(i)} = \partial_\mu A_{\nu\,m}^{(i)} - \partial_\nu
A_{\mu\,m}^{(i)}$, with $A_{\mu\,m}^{(i)}$ being the gauge fields
resulting from the dimensional reduction of the NS-NS and R-R sector
antisymmetric tensor fields. Finally, ${\cal H}_{\mu\nu\lambda} \equiv 
\left(\begin{array}{c} H_{\mu\nu\lambda}^{(1)} \\ H_{\mu\nu\lambda}^
{(2)}\end{array}\right)$, $H_{\mu\nu\lambda}^{(i)} = \left(
\partial_\mu B_{\nu\lambda}^{(i)} - F_{\mu\nu}^{(3)\,m} A_{\lambda\,m}
^{(i)} + {\rm cyc.\,\,in\,\,} \mu\nu\lambda\right)$. The action
(25) can be easily seen to be invariant under the following global
SL(2, R) transformation [13,14]:
\eqabegin
{\cal M} &\rightarrow& \Lambda {\cal M} \Lambda^T, 
\qquad {\cal B}_{mn}
\,\,\,\rightarrow\,\,\,(\Lambda^{-1})^T {\cal B}_{mn}\nn\\ 
\left(\begin{array}{c}  A_{\mu\,m}^{(1)}\\ A_{\mu\,m}^{(2)}\end{array}\right)
&\equiv& {\cal A}_{\mu\,m}\,\,\,\rightarrow\,\,\,(\Lambda^{-1})^T
{\cal A}_{\mu\,m},\qquad
\left(\begin{array}{c} B_{\mu\nu}^{(1)}\\ B_{\mu\nu}^{(2)}\end{array}\right)
\,\,\equiv\,\, {\cal B}_{\mu\nu}\,\,\,\rightarrow\,\,\,(\Lambda^{-1})^T 
{\cal B}_{\mu\nu}\nn\\
 g_{\mu\nu}&\rightarrow& g_{\mu\nu},
\qquad g_{mn}\,\,\,\rightarrow\,\,\,g_{mn},\qquad
{\rm and}\quad A_{\mu}^{(3)\,m}\,\,\,\rightarrow\,\,\,A_\mu^{(3)\,m}
\eqaend
where $\Lambda\,\,=\,\,\left(\begin{array}{cc} a & b\\ c & d\end{array}
\right)$, is the SL(2, R) transformation matrix and $a$, $b$, $c$ and $d$
are constants satisfying $ad -bc =1$.

We shall consider a truncated action, rather than the full action (25).
 Let us, from  now on, set $H_{\mu\nu\lambda}^{(i)} = 0$, $A_\mu^{(3)\,m} = 0$,
$G_{mn} = \delta_{mn}$, $\Delta = 1$, $B_{mn}^{(i)} = 0$ and all the
components of $A_{\mu\,m}^{(1)}$ and $A_{\mu\,m}^{(2)}$ to zero except
one (we call the non-zero components of the gauge fields as $A_\mu^{(1)}$
and $A_\mu^{(2)}$ with the corresponding field-strength $F_{\mu\nu}^{(i)}
= \partial_\mu A_\nu^{(i)} - \partial_\nu A_\mu^{(i)}$), then the action
(25) reduces to:
\eqabegin
& &\int\, d^4x \sqrt{-g}\left[R + \frac{1}{4} {\rm tr}\,\partial_\mu
{\cal M} \partial^\mu {\cal M}^{-1} + \frac{1}{8} \partial_\mu
\log \b {\Delta} \partial^\mu \log \b {\Delta}\right. \nn\\
& &\qquad\qquad\qquad \left. +\frac{1}{4} \partial_\mu g_{mn} \partial^\mu
g^{mn} - \frac{1}{4} (\b {\Delta})^{\frac{1}{6}} {\cal F}_{\mu\nu}^T
{\cal M} {\cal F}^{\mu\nu}\right]
\eqaend
Here ${\cal M}$ is as given in (26) with $\tilde {\phi}$ replaced 
by $\phi$ since we have set $\Delta = 1$. ${\cal F}_{\mu\nu} \equiv
\left(\begin{array}{c} F_{\mu\nu}^{(1)} \\ F_{\mu\nu}^{(2)}\end{array}
\right)$. The action (28) is invariant under the global SL(2, R)
transformation:
\eqabegin
{\cal M} &\rightarrow& \Lambda {\cal M} \Lambda^T, \qquad
\left(\begin{array}{c} A_\mu^{(1)} \\ A_\mu^{(2)}\end{array}\right)
\,\,\,\equiv\,\,\, {\cal A}_\mu \,\,\,\rightarrow \,\,\, (\Lambda^{-1})^T
{\cal A}_\mu\nn\\
g_{\mu\nu} &\rightarrow & g_{\mu\nu} \qquad {\rm and} \qquad g_{mn}\,\,\,
\rightarrow \,\,\, g_{mn}
\eqaend
Note  that type IIB string effective action (28) reduces
precisely to the action (3) considered by GHS, when the R-R fields are
set to zero. We would like to exploit this SL(2, R) invariance to rotate
the magnetically charged black hole solution of GHS to a more general
black hole solution of type II string theory in $D = 4$. In order to 
describe the complete black hole solution, we have to specify the asymptotic
values of the dilaton $\phi$ and R-R scalar $\chi$. Under the transformation
(29), the complex scalar field $\lambda = \chi + i e^{-\phi}$ and the gauge
field $A_\mu^{(i)}$ transform as follows,
\eqabegin
\lambda &\rightarrow& \frac{a\lambda + b}{c\lambda + d}\\
A_3^{(1)} &\rightarrow& d A_3^{(1)} - c A_3^{(2)}\nn\\
A_3^{(2)} &\rightarrow & -b A_3^{(1)} + a A_3^{(2)}
\eqaend
Note from (9) that the only non-zero component of the field-strength
$F_{\mu\nu}^{(1)}$ is $F_{23}^{(1)} = Q \sin\theta$ and so, upto the gauge
transformation the only non-zero component of the gauge field $A_\mu^{(1)}$
is $A_3^{(1)} = - Q\cos\theta$. Let us first construct the black hole 
solution for the simplest choice of $\lambda_0 = i$ (i.e. for $\phi_0 = 
\chi_0 = 0$), where the subscript `zero' represent the 
asymptotic value of scalars in ${\cal M}$ and this black hole carries 
only NS-NS charge.
 Here $Q$ can
be argued to be quantized in some basic units. Although the actions (25) and
(28) are invariant under SL(2,R) transformations, in the quantized theory,
the remnant, robust symmetry is expected to be SL(2,Z) and elements of $
\Lambda $ are integers satisfying the constraint 
det $\Lambda =1$. Thus starting
from a black hole with a given $Q$ and 
$\lambda _0 =i$, we can obtain a black 
hole which carries both type of charges. The relevant transformation matrix
has the following form:
\begineq
\Lambda = \frac{1}{\sqrt{q_1^2 + q_2^2}}\left(\begin{array}{cc}
q_1 & -q_2\\ q_2 & q_1\end{array}\right)
\endeq
where $q_1$ and $q_2$ are two integers and still $\lambda _0 = i$. 
Note here that we have used the fact
that the charges $\left(\begin{array}{c} Q^{(1)} \\ Q^{(2)}
\end{array}\right)$ associated with the gauge fields $A_\mu^{(1)}$ and
$A_\mu^{(2)}$ transform as $\Lambda \left(\begin{array}{c} Q^{(1)} 
\\ Q^{(2)}\end{array}\right)$ eventhough the gauge fields themselves
transform as given in (29). This is in contrary to the usual Maxwell
theory where both the gauge fields and the charges should transform
in the same way. This difference can be understood by looking at the 
gauge field kinetic term in our action (28). The equation of motion
in this case has the form
\begineq
\nabla_\mu\left({\cal M} {\cal F}^{\mu\nu}\right) = {\cal J}^\nu
\endeq
It is clear from (33) that the charges would transform contragradiently
with respect to the gauge fields 
under the SL(2, Z) transformation [18]
as happened in our case. Once we have derived the form of $\Lambda$, we can
easily calculate the gauge field components and the complex scalar
from (31) and (30) as follows:
\eqabegin
A_3^{(i)} &=& - q_i Q \cos\theta\\
\lambda &=& \frac{i q_1 R\left(1 - f^2\right) - 2 q_2 M}{i q_2 R\left(1 -
f^2\right) + 2 q_1 M}\nn\\
&=& \frac{q_1 q_2 \left[R^2\left(1 - f^2\right)^2 - 4M^2\right] +
2iMR\left(1 - f^2\right)\left(q_1^2 + q_2^2\right)}{4 q_1^2 M^2 + q_2^2
R^2 \left(1 - f^2\right)^2}
\eqaend
We note from (35) that asymptotically $R\left(1 - f^2\right) \rightarrow
2M$ and therfore $\lambda \rightarrow i$ as $r \rightarrow \infty$.

Let us now generalize our construction for an arbitrary vacuum modulus 
$\lambda_0$. In this case we replace the charge $Q$ with an arbitrary
value $\alpha_{(q_1, q_2)} = \Delta^{1/2}_{(q_1, q_2)} Q$. $\Delta^{1/2}_
{(q_1, q_2)}$ will be determined later. We take the SL(2, R) transformation
matrix as
\eqabegin
\Lambda\,\,\,=\,\,\,\Lambda_1 \Lambda_2 &=& \left(\begin{array}{cc}
e^{-\phi_0/2} & \chi_0 e^{\phi_0/2} \\ 0 & e^{\phi_0/2}\end{array}\right)
\left(\begin{array}{cc} \cos\alpha & -\sin\alpha\\ \sin\alpha & \cos\alpha
\end{array}\right)\nn\\
&=&\left(\begin{array}{cc} e^{-\phi_0} \cos\alpha + \chi_0 \sin\alpha &
-e^{-\phi_0} \sin\alpha + \chi_0 \cos\alpha\\ \sin\alpha & \cos\alpha
\end{array}\right) e^{\phi_0/2}
\eqaend
Here $\Lambda_2$ is the most general SL(2, R) matrix which preserves
the vacuum modulus $\lambda_0 = i$ and $\Lambda_1$ is the SL(2, R)
matrix which transforms it to an arbitrary value $\lambda = \lambda_0$.
$\alpha$ is an arbitrary parameter which will be fixed from the
charge quantization condition. Now since we have $\Lambda$, we can
find the magnetic charges associated with the gauge fields $A_3^{(1)}$
and $A_3^{(2)}$ as,
\eqabegin
Q^{(1)} &=& \left(e^{-\phi_0/2} \cos\alpha + \chi_0 e^{\phi_0/2} \sin\alpha
\right) \Delta_{(q_1, q_2)}^{1/2} Q\nn\\
Q^{(2)} &=& e^{\phi_0/2} \sin\alpha \Delta_{(q_1, q_2)}^{1/2} Q
\eqaend
Using the charge quantization condition, we get from (37),
\eqabegin
\sin\alpha &=& e^{-\phi_0/2} \Delta^{-1/2}_{(q_1, q_2)} q_2\nn\\
\cos\alpha &=& e^{\phi_0/2} \left(q_1 - q_2 \chi_0\right) \Delta^{-1/2}_
{(q_1, q_2)}
\eqaend
where $q_1$, $q_2$ are integers. $\Delta_{(q_1, q_2)}$ can be evaluated
if we use $\sin^2 \alpha + \cos^2 \alpha = 1$, as follows,
\eqabegin
\Delta_{(q_1, q_2)} &=& e^{-\phi_0} q_2^2 + \left(q_1 - q_2 \chi_0
\right)^2 e^{\phi_0}\nn\\
&=& (q_1, q_2) {\cal M}_0^{-1} \left(\begin{array}{c} q_1 \\ q_2\end{array}
\right)
\eqaend
where ${\cal M}_0 = \left(\begin{array}{cc}\chi_0^2 + e^{-2\phi_0} &
\chi_0\\ \chi_0 & 1\end{array}\right)e^{\phi_0}$. We note that the 
expression for $\Delta_{(q_1, q_2)}$ is SL(2, Z) invariant and therfore
the charges of the black holes $\alpha_{(q_1, q_2)} = \Delta^{1/2}_{(q_1,
q_2)} Q$ are also given by SL(2, Z) covariant expressions. With the 
$\Lambda$ in (36), we write below the transformed gauge fields,
\eqabegin
A_3^{(1)} &=& - e^{-\phi_0} \left(q_1 - q_2\chi_0\right) Q \cos\theta\nn\\
A_3^{(2)} &=& - e^{-\phi_0} \left(q_2 |\lambda|^2 - q_1 \chi_0\right) Q
\cos\theta
\eqaend
which can be written compactly as,
\begineq
\left(\begin{array}{c} A_3^{(1)} \\ A_3^{(2)} \end{array}\right) = 
- {\cal M}_0^{-1} \left(\begin{array}{c} q_1 \\ q_2\end{array}\right)
Q \cos\theta
\endeq
Also, the value of the transformed complex scalar field is:
\eqabegin
\lambda &=& \frac{i q_1 e^{-\phi_0} R \left(1 - f^2\right) + 2M \left
(q_1 \chi_0 - q_2 |\lambda_0|^2\right)}{ i q_2 e^{-\phi_0} R\left(
1 - f^2\right) + 2M \left(q_1 - q_2\chi_0\right)}\nn\\
&=&\frac{4M^2 e^{-\phi_0} \chi_0 \Delta_{(q_1, q_2)} + q_1 q_2 e^
{-2\phi_0}\left(R^2\left(1 - f^2\right)^2 - 4M^2\right) + i 2MR \left(
1 - f^2\right) e^{-2\phi_0} \Delta_{(q_1, q_2)}}
{4M^2\left(q_1 - q_2 \chi_0
\right)^2 + q_2^2 e^{-2\phi_0} R^2 \left(1 - f^2\right)^2}\nn
\\
\eqaend
It can be checked that asymptotically as $r \rightarrow \infty$,
$R(1 - f^2) \rightarrow 2M$ and so, $\lambda \rightarrow \lambda_0$. From
(42) we obtain
\begineq
e^{-2\phi} = \frac{4M^2 R^2 \left(1 - f^2\right)^2 e^{-4\phi_0}
\Delta^2_{(q_1, q_2)}}{\left[4M^2\left(q_1 - q_2\chi_0\right)^2
+q_2^2 e^{-2\phi_0} R^2\left(1 - f^2\right)^2\right]^2}
\endeq
So, starting from the magnetically charged black hole solution of
GHS, we have constructed an SL(2, Z) multiplet of magnetically 
charged black hole solution of type II string theory given by
the field configurations (19) and (40--43). Note that in (19)
$Q$ is now replaced by $\Delta^{1/2}_{(q_1, q_2)} Q$. The black hole
solutions in this case are characterized by two integers
corresponding to the magnetic charges associated with the two
gauge fields of the NS-NS and R-R sectors. Since the canonical
metric does not transform under the SL(2, R) transformation the
transition between the black hole and the ``naked'' singularity
occurs at the same point ($Q^2 = 8M^2 e^{-2\phi_0}$) as discussed 
earlier. However, we can no longer express the metric
nicely in terms of the new dilaton charge as was done in Eq.(22).
Therefore, we note that
in the extremal case, mass of the black hole should also be
given by $M_{(q_1, q_2)} = \Delta^{1/2}_{(q_1, q_2)} M$ and
the black holes will be BPS saturated.

Since
\begineq
M_{(q_1, q_2)} = \sqrt{e^{-\phi_0} q_2^2 + \left(q_1 - q_2 
\chi_0\right)^2 e^{\phi_0}} M
\endeq
masses satisfy the following relation when $\chi = 0$,
\eqabegin
& &\left(M_{(q_1, q_2)} + M_{(p_1, p_2)}\right)^2 - M_{(q_1 + p_1, q_2 
+ p_2)}^2\nn\\
&=& 2M^2\left\{\left[\left(p_1 q_1 e^{\phi_0} + p_2 q_2 e^{-\phi_0}
\right)^2 + \left(p_1 q_2 - p_2 q_1\right)^2\right]^{1/2}
-\left(p_1 q_1 e^{\phi_0} + p_2 q_2 e^{-\phi_0}\right)\right\}\nn\\
&\geq& 0
\eqaend
As $M$ is a real, positive number (45) gives a triangle inequality
among the masses which we write below:
\begineq
M_{(q_1, q_2)} + M_{(p_1, p_2)} \geq M_{(q_1 + p_1, q_2 + p_2)}
\endeq
The equality holds when $p_1 q_2 = p_2 q_1$ i.e. when $p_1 = n q_1$
and $p_2 = n q_2$, with $n$ being an integer. Therefore, when $q_1$,
$q_2$ are relatively prime integers, the inequality prevents the black
holes to decay into black holes of lower masses. So, because of the
`mass gap' relation (46), the extremal black holes are  stable.
Furthermore, note that since charges also satisfy the same relation
(44) like the masses, when $q_1$ and $q_2$ are relatively prime the
charge conservation can not be satisfied if the black holes decay [25].
Finally, we note from (19) that even in the case of type II black
hole solution the area of the event horizon vanishes in the extremal
limit, like what happened for the heterotic string case of GHS, causing
the surface to be singular.




To summarize, we  first argued that  the magnetically charged black 
hole solution GHS derived in the context of $D=4$ heterotic string theory 
can also be interpreted as black hole solution of $D=4$ type IIB theory such
that the gauge field appears due to compactification of the NS-NS antisymmetric
field (of $D=10$ action) with all R-R fields set to zero. 
It has been demonstrated
 that
 the low energy effective action  of type IIB
string theory compactified on torus  possesses
an SL(2, Z) invariance if the $D=10$ theory is endowed with the same symmetry.
 By exploiting  this symmetry of type IIB string theory, we have 
constructed an infinite family of magnetically charged black hole solutions
in $D = 4$.
Black hole solutions in string theory having electric, magnetic
and both charges associated with the gauge fields originating from the
dimensional reduction of the various heterotic string states 
as well as the NS-NS sector states of type II string theory have been 
constructed before. The solutions we have constructed in this paper
are characterized by two integers corresponding to the charges
associated with both NS-NS sector and R-R sector gauge fields. We
have shown that in the extremal limit, when these two integers 
$(q_1, q_2)$  
are relatively prime, the black holes are  stable as
they are prevented from decaying due to the inequality (46). In this context,
we are tempted to interprete the BPS saturated $(q_1, q_2)$ black 
holes as marginal
bound states of $q_1$ NS-NS black holes and $q_2$ R-R black holes 
analogous to
the arguments due to Witten [26] for the string solutions of Schwarz [18]. 
The class of
black hole solutions obtained here have zero area of event horizon. It is well
known that in order to construct $D=4$ extremal black holes, with nonzero area,
one must have four nonzero charges (corresponding to the number of
1-D-branes, the number of 5-D-branes, the number of solitonic 5-branes
or Kaluza-Klein monopoles and the Kaluza-Klein charges) 
as has been discussed by several authors [12]
in the context of intersecting D-brane approach [27]. 
In our case, the solutions are
characterized by only two charges. However, the computation of entropy for
the type of black holes, presented here, can be carried out by using the 
concept of ``stretched horizon'' [28].  
 It will also be interesting to construct other
classes of black hole solutions in type IIB string theory by implementing
the electric-magnetic duality transformations,
for example, the dyonic black holes for which the area is known
not to vanish.

\vspace{1cm}

\begin{large}
\noindent{\bf Acknowledgements}
\end{large}

\vspace{.5cm}

A.D. would like to thank the members of the theory group at Saha Institute
of Nuclear Physics, where this work was carried out, for hospitality. This
work is supported in part by US DOE Grant no. DE-FG-02-91ER40685.

\vspace{1cm}

\begin{large}
\noindent{\bf References}
\end{large}

\vspace{.5cm}

\baselineskip=12pt
\begin{enumerate}
\item R. C. Myers and M. J. Perry, Ann. Phys. 172 (1986) 304; R. C.
Myers, \np 289 (1987) 701; C. G. Callan, R. C. Myers and M. J. Perry,
\np 311 (1988) 673; H. J. de Vega and N. Sanchez, \np 309 (1988) 552;
P. O. Mazur, Gen. Rel. Grav. 19 (1987) 1173.
\item S. W. Hawking, Comm. Math. Phys. 43 (1975) 199; Phys. Rev. D14
(1976) 2460; R. D. Carlitz and R. S. Willey, Phys. Rev. D36 (1987) 2327,
2336; S. B. Giddings, Phys. Rev. D46 (1992) 1347; J. Preskill,
{\it Do Black Holes Destroy Information?}, hep-th/9209058; D. N. Page,
{\it Black Hole Information}, hep-th/9305040.
\item G. Gibbons, \np 207 (1982) 337; G. Gibbons and K. Maeda, \np
298 (1988) 741.
\item D. Garfinkle, G. Horowitz and A. Strominger, Phys. Rev. D43 (1991)
3140; Erratum: Phys. Rev. D45 (1992) 3888.
\item J. Preskill, P. Schwarz, A. Shapere, S. Trivedi and F. Wilczek,
Mod. Phys. Lett. A6 (1991) 2353; A. Shapere, S. Trivedi and F. Wilczek,
Mod. Phys. Lett. A6 (1991) 2677; C. F. E. Holzhey and F. Wilczek, \np
380 (1992) 447.
\item R. Kallosh, \pl 282 (1992) 80; R. Kallosh, A. Linde, T. Ortin, 
A. Peet and A. van Proeyen, Phys. Rev. D46 (1992) 5278.
\item A. Sen, Phys. Rev. Lett. 69 (1992) 1006; \np 440 (1995) 421.
\item M. Cvetic and D. Youm, Phys. Rev. D53 (1996) 584; D. Jatkar,
S. Mukherji and S. Panda, \np 484 (1997) 223.
\item A. Strominger and C. Vafa, \pl 379 (1996) 99.
\item J. Polchinski, Phys. Rev. Lett. 75 (1995) 4724; J. Polchinski,
S. Chaudhuri and C. V. Johnson, {\it Notes on D-Branes}, 
hep-th/9602052; J. Polchinski, {\it Lectures on D-Branes}, 
hep-th/9611050.
\item C. Callan and J. Maldacena, \np 475 (1996) 645; G. Horowitz and
A. Strominger, Phys. Rev. Lett. 77 (1996) 2368; G. Horowitz, J. Maldacena
and A. Strominger, \pl 383 (1996) 151; S. Das and S. Mathur, {\it 
Comparing Decay Rates for Black Holes and D-Branes}, hep-th/9608185;
S. Gubser and I. Klebanov, {\it Emission of Charged Particles from
Four- and Five-Dimensional Black Holes}, hep-th/9608108.
\item M. Cvetic and A. Tseytlin, \pl 366 (1996) 95;
 J. Maldacena and A. Strominger, Phys. Rev. Lett. 77 (1996) 428;
C. Johnson, R. Khuri and R. Myers, \pl 378 (1996) 78.
\item J. Maharana, Phys. Lett. B402 (1997) 64. 
\item S. Roy, {\it On S-Duality of Toroidally Compactified Type IIB
String Effective Action}, hep-th/9705016.
\item B. Julia, in {\it Supergravity and Superspace}, Eds. S. W. 
Hawking and M. Rocek, Cambridge University Press, 1981.
\item C. M. Hull and P. K. Townsend, \np 438 (1995) 109.
\item C. Hull, \pl 357 (1995) 545.
\item J. H. Schwarz, \pl 360, (1995) 13 (hep-th/9508143, see  revised 
version, June, '97); {\it Superstring Dualities}, hep-th/9509148.
\item S. Roy, {\it SL(2, Z) Multiplets of Type II Superstrings in 
$D < 10$}, hep-th/9706165.
\item E. Cremmer and B. Julia, \np 159 (1979) 141.
\item A. Sen, Int. Jour. Mod. Phys. A9 (1994) 3707; M. J. Duff, R. R. Khuri
and J. X. Lu, Phys. Rep. C259 (1995) 213.
\item A. Giveon, M. Porrati and E. Rabinovici, Phys. Rep. C244 (1994) 77;
E. Alvarez, L. Alvarez-Gaume and Y. Lozano, {\it An Introduction to T-Duality
in String Theory}, hep-th/9410237.
\item J. Maharana and J. Schwarz, \np 390 (1993) 3; S. F. Hassan and A. Sen,
\np 375 (1992) 103.
\item N. Marcus and J. H. Schwarz, \pl 115 (1982) 111; J. H. Schwarz,
\np 226 (1983) 269; P. Howe and P. West, \np 238 (1984) 181.
\item A similar discussion in the context of string solution is given
in J. H. Schwarz, {\it Lectures on Superstring and M-Theory Dualities},
hep-th/9607201.
\item E. Witten, \np 460 (1996) 335.
\item A. Tseytlin, \np 475 (1996) 149; I. Klebanov and A. Tseytlin,
\np 475 (1996) 179; J. Gauntlett, D. Kastor and J. Traschen, {\it
Overlapping Branes in M-Theory}, hep-th/9604179, S. Ferrara and J. Maldacena,
hep-th/9706097.
\item L. Susskind, {\it Some Speculations about Black Hole
Entropy in String Theory}, hep-th/9309145; L. Susskind and J. Uglum,
Phys. Rev. D50 (1994) 2700; L. Susskind, Phys. Rev. D49 (1994) 6606;
A. Sen, Mod. Phys. Lett. A10 (1995) 2081.
\end{enumerate}

\vfil
\eject
\end{document}